\RequirePackage{lineno}
\documentclass{nature}
\usepackage{xcolor}
\usepackage{amsmath}
\usepackage{caption}
\usepackage{graphicx}
\usepackage{tabularx}
\usepackage{makecell}
\makeatletter
\let\saved@includegraphics\includegraphics
\AtBeginDocument{\let\includegraphics\saved@includegraphics}
\renewenvironment*{figure}{\@float{figure}}{\end@float}
\makeatother
\usepackage{ctex,multicol,multirow,graphics,fancyhdr,epstopdf}
\AtBeginDocument{}
\usepackage{amsmath,amsfonts,amssymb,bm,upgreek,mathrsfs,ccmap,mathcomp}

\usepackage{amsmath, amssymb, amsfonts}
\usepackage{bm}
\usepackage{siunitx}
\usepackage{mhchem}
\usepackage{graphicx}
\usepackage{hyperref}

\title{Robust Floquet-induced gap in irradiated graphite}

\author{Fei Wang$^{1,2,\ast}$, Xuanxi Cai$^{1,2,\ast}$, Wanying Chen$^{1,2,\ast}$, Jinxi Lu$^{1,2}$, Tianshuang Sheng$^{1,2}$, Xiao Tang$^{1,2}$, Jiansong Li$^{1,2}$, Hongyun Zhang$^{1,2}$ \& Shuyun Zhou$^{1,2,3,\dagger}$}

\usepackage{graphicx}
\makeatletter
\let\saved@includegraphics\includegraphics
\AtBeginDocument{\let\includegraphics\saved@includegraphics}

\makeatother

\begin{document}
\maketitle

\begin{affiliations}

\item Department of Physics, Tsinghua University, Beijing 100084, People's Republic of China
\item State Key Laboratory of Low-Dimensional Quantum Physics, Tsinghua University, Beijing 100084, People's Republic of China
\item Frontier Science Center for Quantum Information, Beijing 100084, People's Republic of China

* These authors contributed equally to this work\\
$\dagger$ Correspondence and request for materials should be sent to syzhou@mail.tsinghua.edu.cn

\end{affiliations}

\begin{abstract}

Floquet engineering provides an emerging pathway for tailoring the electronic states of quantum materials through time-periodic drive\cite{oka2009prb}. A critical step along this direction is achieving light-induced modifications of the dynamical electronic structure, such as avoided-crossing gap at the Floquet Brillouin zone boundary, via efficient coupling of electrons with the coherent light-field. Here, we report robust Floquet-induced gap in bulk graphite that persists despite the presence of interlayer coupling and photo-excitation. Using time- and angle-resolved photoemission spectroscopy with intense mid-infrared pumping, we directly reveal Floquet-induced gaps at resonance points both in the valence and conduction bands, accompanied by coherent Floquet sidebands. The gap and sidebands coexist with photo-excited carriers, yet their distinct timescales allow us to disentangle their origins. Our demonstration of robust Floquet-induced gaps establishes graphite as a platform for coherent manipulation of Dirac fermions and realization of light-engineered quantum phases.

\end{abstract}

\section*{Introduction}

Floquet engineering provides a powerful framework for designing non-equilibrium phases of matter with tailored electronic structures and emergent properties through time-periodic driving. In analogy to Bloch states, where a spatially-periodic potential leads to band folding in the momentum Brillouin zone (BZ), time-periodic light-field can result in band folding in the energy BZ. More importantly, interaction of the electronic states with the periodic potential could lead to hybridization (avoided-crossing) gap opening at the BZ boundary and symmetry breaking, enabling light-induced control of the electronic structure and material properties\cite{HsiehNM2017,oka2019floquet,SentefRMP2021,ZhouNRP2021}.
 
As a two-dimensional material hosting characteristic Dirac cone, graphene provides a model system for Floquet engineering, where light-induced anomalous Hall effect was predicted by breaking the time-reversal symmetry with circularly polarized light\cite{oka2009prb}.
Despite extensive theoretical predictions\cite{EfetovPRB2008,NaumisPRB2008,NaumisPhilosMag2010,AuerbachGRPRL2011,AlexandrePRB2011,TorresAPL2011,wu2011prb,PlateroPRB2013,Balseiro2014PRB,SavelevTimeCrysPRB2014,Devereaux2015NC,MengSGPRB2019,LudwigPRR2022}, an intrinsic difficulty is that the drive is often accompanied by photo-excitation across the Dirac cone. Such photo-excitation triggers many-body scatterings, such as electron-electron (e-e), electron-phonon (e-ph) and phonon-phonon (ph-ph) interactions, that can disrupt the coherence of Floquet engineering\cite{Cavalleri2013NM,gierz2015tracking,Damascelli2019science,GierzNanoLett2021,ZhouNSR2021,wu2024ultrafast,wu2020ultrafast}. Over the past decade, Floquet-induced gaps have been experimentally observed in topological insulators\cite{Gedik2013,Gedik2016,Mahmood2025floquet} and black phosphorus\cite{zhou2023pseudospin}. In graphene, Floquet-Bloch states manifested through Floquet-Andreev states\cite{lee2022nat} and Floquet-Volkov interference\cite{GierzNanoLett2021,gedik2025graphene,stefan2025graphene} have been reported recently. Moreover, Floquet-induced gap has been achieved in epitaxial graphene\cite{wang2026observation}, where the sample is strongly electron doped such that direct photo-excitation from the lower Dirac cone to upper Dirac cone is blocked. For charge neutral graphene sample, photo-excitation across the Dirac point is allowed, whether the gap can survive in the presence of photo-excitation and subsequent scattering is a fundamental question that awaits to be addressed.

Moving from graphene to its three-dimensional counterpart, graphite, the interlayer interaction modifies the electronic structure\cite{ZhouNP2006}, resulting in coexistence of massless Dirac cone near the H point ($k_z = \pi/c$, where $c$ is the out-of-plane lattice constant), and splitting bands with parabolic touching near the K point ($k_z = 0$) [Fig.~1(a)]. While Floquet engineering of graphite have been theoretically proposed,\cite{gupta2003generation,oka2011all,graphite2020prb,yaoyugui2018multilayer} experimental progress has been lacking. For charge-neutral graphite, where $E_F$ lies near the Dirac point, a mid-infrared (MIR) pump pulse can play dual roles: dressing the electronic states via time-periodic light-field while simultaneously photo-exciting electrons from the lower to upper Dirac cone, similar to hole-doped Bi$_2$Te$_3$\cite{WFBi2Te32025,chen2024distinct}. The presence of interlayer coupling and additional bands in graphite also introduces extra scattering channels, which could, in principle, be detrimental to coherent light-field dressing or even smear out the gap features. This raises a fundamental question about Floquet engineering and electron scattering: can light-induced avoided-crossing gaps persist in the presence of interlayer coupling and its associated scattering channels in graphite?

Here, we provide direct experimental evidence for the survival of Floquet-induced gaps in graphite despite the interlayer coupling and additional scattering channels in graphene. Using time- and angle-resolved photoemission spectroscopy (TrARPES) combined with intense MIR pumping [Fig.~1(a)], we observe clear Floquet-induced gaps in both the valence band (VB) and the photo-excited conduction band (CB) above the Fermi level ($E_F$), as schematically illustrated in Fig.~1(b) and (c). These gaps emerge concurrently with Floquet sidebands and the driving field, observed on timescales much shorter than carrier relaxation. The observation of Floquet-induced gaps in graphite demonstrates the robustness of light-field driven band engineering in the presence of interlayer coupling and scatterings, establishing bulk graphite as a promising platform for realizing light-engineered quantum phases.

\begin{figure*}[htbp]
	\centering
	\includegraphics[width=16.8cm]{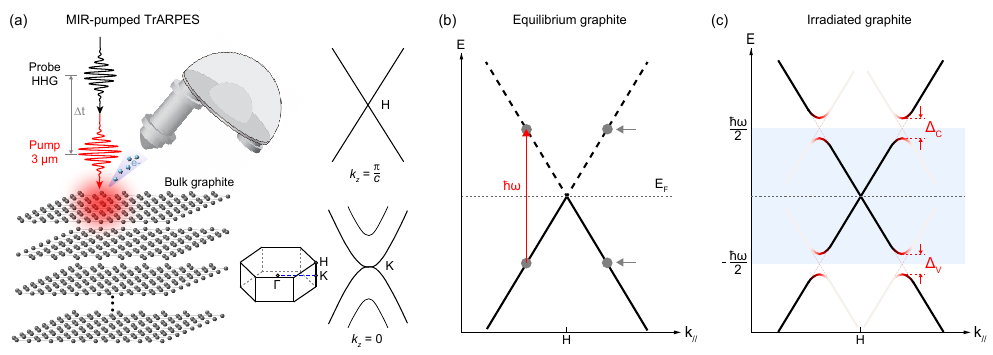}
	\caption*{Fig.~1. Schematic illustration for light-induced avoided-crossing gap in Dirac cone of graphene/graphite. (a) Schematic of MIR pumped TrARPES experiment. The inset shows the Brillouin zone of graphite and dispersions near H and K. (b,c) Conical dispersion in the equilibrium state (b) and light-induced gap opening at the Floquet resonance points upon resonant driving with $\hbar\omega$ (c).}
\label{Fig1}
\end{figure*}

\section*{Methods}

The sample is an exfoliated graphite flake, with a high-quality, flat region of approximately 1 mm in size [see the optical image in the inset of Fig.~2(a)]. Such large sample  area with reduced scattering allows to impinge a strong laser pulse at a large pump fluence of 5.2 mJ/cm$^2$, which is critical for achieving strong light-matter coupling.  TrARPES measurements were performed under a base pressure better than 3.6 $\times$ 10$^{-11}$ Torr at 80 K. The probe beam is a high harmonic generation (HHG) light source with a photon energy of 21.7 eV, pulse duration of 66 fs and a flux of 3$\times$10$^{11}$ photons/s\cite{syzhou2024hhg}. The pump beam, with wavelength $\lambda$ = 3 $\mu$m corresponding to photon energy of $\hbar\omega$ = 415 meV, was generated via non-collinear difference frequency generation (NDFG) between the signal and idler beams from an optical parametric amplifier (OPA). The pump beam was $p$-polarized and aligned perpendicular to the measured $k_{y}$ direction of the sample, while the probe beam was $s$-polarized (parallel to $k_{y}$). The  low pump photon energy was chosen such that the first-order sidebands ($n = \pm 1$) of the Dirac cone overlap with the equilibrium Dirac cone, at momentum points $k$ that satisfy the resonance condition, $E_{CB,k}-E_{VB,k}=\hbar\omega$. 

\section*{Results and Discussions}

Figure 2(a) shows dispersion image measured through the Brillouin zone corner at $\Delta t$ = -200 fs before pumping, along the direction marked by the red line in the inset. ARPES measurements have shown that the electronic structure of graphite shows conical dispersion  near the H point similar to monolayer graphene, while near the K point, two splitting bands are observed\cite{ZhouNP2006}. Here a clear conical dispersion similar to that in monolayer graphene is observed, indicating that we are probing the Dirac cone near the H point at the probe photon energy.  

Figures 2(b)-2(g) show snapshots of dispersion images measured at different delay times upon pumping with $\hbar\omega$ = 415 meV at pump fluence of 5.2 mJ/cm$^2$. The corresponding differential images (obtained by subtracting the image at -200 fs) are shown in Figs.~2(h)-2(m). 
 Upon pumping, the electronic structure exhibits pronounced modifications. These include not only the photo-excitation from the VB to the CB, but also, more importantly, by a strong suppression of the VB intensity at the resonance points in the VB and CB [indicated by red arrows in Fig.~2(d)] and the emergence of Floquet sidebands. These sidebands appear as replicas of the Dirac cone shifted by the pump photon energy and are more clearly visible in the differential images shown in Figs.~2(i)-2(k).

Since photo-excitation and light-field dressing coexist upon pumping the Dirac cone, we  disentangle their contributions through detailed analysis of their dynamical evolution. Figure 2(n) shows the time evolution of the integrated intensity for the photo-excited conduction band [P1 marked in Fig.~2(c)], Floquet sidebands [S2 and S3 in Fig.~2(c)], and the resonance point [G4 in Fig.~2(c)]. Following photo-excitation, the CB intensity exhibits a rising edge that peaks shortly after time zero, followed by slow carrier relaxation over several hundred femtoseconds [see P1 in Fig.~2(n)]. Note that the relaxation of photo-excited electrons depends on their energy; for electrons near $E_F$, the relaxation time can extend to a few picoseconds, as revealed by time-resolved photoemission electron microscopy\cite{zhang2025spatial}. In contrast, the light-induced sidebands at S3, which represent the convolution of the pump and probe pulses, are observed only near time zero, reaching maximum intensity at time zero [see S3 in Fig.~2(n)]. This shows that photo-excited carrier scattering occurs on a much longer timescale than the formation of Floquet states. 
 
 With the understanding of the distinctive timescales for carrier relaxation and Floquet states, we now analyze the dynamical suppression of the VB intensity near time zero. First, we examine the sideband above $E_F$ [labeled as S2 in Fig.~1(c)], which involves both light-field dressing and  carrier scattering from the CB. As shown in the time trace in Fig.~2(n), its dynamics  can be fitted with two components: an exponential decay representing carrier relaxation over a few hundred femtosecond (gray fitting curve), and a Gaussian peak representing the Floquet sideband dynamics (black fitting curve), similar to that observed at S3. Interestingly, the strong suppression of the VB intensity at the resonance point G4 also exhibits two components: one following the Floquet state dynamics [red dotted curve in Fig.~2(n)] and another tracking the photo-excitation dynamics (gray curve). This suggests that the suppression of VB intensity at the resonance points $k$ at time zero is not solely due to charge depletion, i.e., photo-excitation of electrons from the VB to the CB leaving holes behind in the VB, but rather includes another component associated with light-induced gap opening as schematically illustrated in Fig.~2(o).

\begin{figure*}[htbp]
	\centering
	\includegraphics[width=16.8cm]{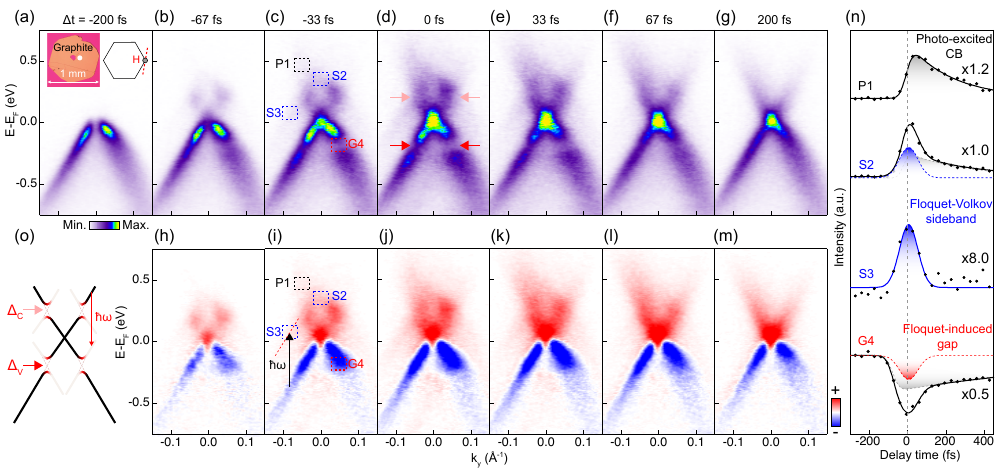}
	\caption*{Fig.~2. Observation of light-induced gap, Floquet-Volkov sideband, photo-excited carriers and their dynamical evolution upon pumping.  (a)-(g) Dispersion images measured at different delay times upon pumping with photon energy $\hbar\omega$ = 415 meV at pump fluence of 5.2 mJ/cm$^2$. The inset shows the optical image of the graphite sample and the Brillouin zone. The white dot in the optical image represents the probe beam spot size. (h)-(m) Differential image obtained by subtracting data measured at $\Delta$t = -200 fs (a) from data in (b)-(g). (n) Integrated intensity of the photo-excited CB (P1), Floquet-Volkov sideband (S2, S3) and VB (G4) as a function of delay time over regions marked in (c) and (i). (o) A schematic summary of the observed dispersion upon pumping.
	}\label{Fig2}
\end{figure*}

Figure 3 shows a detailed analysis of data measured at $\Delta t$ = 0 and $\Delta t$ = 200 fs, revealing the gap opening and enabling extraction of the gap size. The dispersion image at 200 fs [Fig.~3(d)], where the upper Dirac cone is transiently populated, provides a reference for the equilibrium Dirac cone dispersion. The conical dispersion can be extracted by tracking the positions of the two dispersing peaks in Fig.~3(d). At $\Delta t$ = 0 fs, the dispersion near the resonance points is strongly modified by the light-field, as evident in the energy distribution curve (EDC) analysis in Fig.~3(c). While EDCs away from the resonance points remain unchanged [black curves in Fig.~3(c)], those near the resonance points (\textit{k}$_3$, \textit{k}$_4$ and \textit{k}$_8$, \textit{k}$_9$, red curves) exhibit a distinct multiple-peak structure. Zoomed-in EDC analysis at resonance point $k_3$ shows that upon pumping, each of these two peaks, corresponding to VB [Fig.~3(g)] and CB [Fig.~3(h)] respectively, clearly splits into two peaks in Figs.~3(e)-3(f), signaling gap opening at the resonance point for both VB and CB. From the peak separation in the EDC analysis, the light-induced gap is extracted to be $\Delta$ = 158 $\pm$ 15 meV from $k_3$. The gap extracted from the resonance point on the right-hand side at $k_9$ is 220 meV. The observed gap is overall similar to that observed in monolayer graphene at similar pump fluence\cite{wang2026observation}. The extracted electronic structures are plotted in Figs.~3(i)-3(j), showing the emergence of light-induced hybridization gap at $\Delta t$ = 0 and recovery to the Dirac cone at $\Delta t$ = 200 fs. 

\begin{figure*}[htbp]
	\centering
	\includegraphics[width=16.8cm]{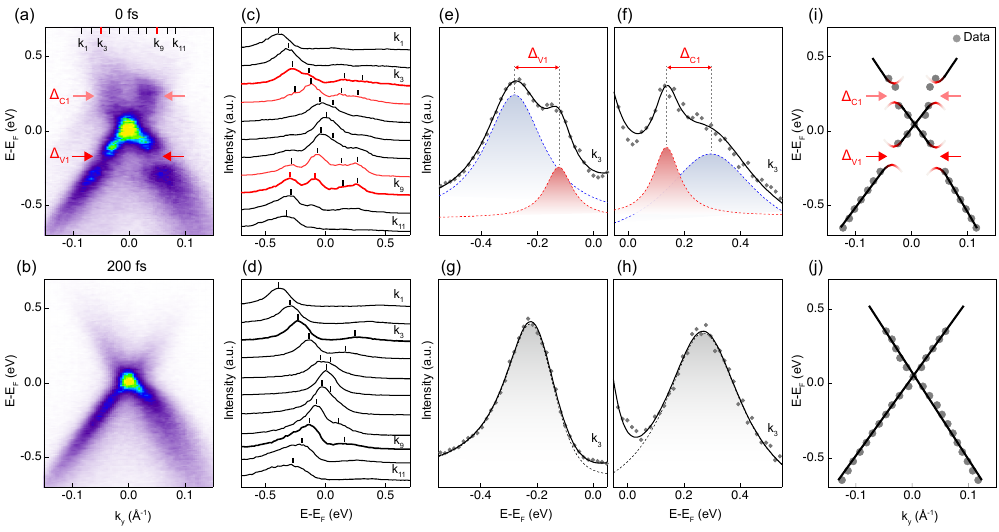}
	\caption*{Fig.~3. Observation of light-induced gap in the conduction band and valence band, and extraction of the gap size from EDC analysis. (a)-(b) Dispersion image measured upon pumping (a) and at 200 fs after pumping (b). (c)-(d) EDCs extracted from (a) and (b)respectively. Tick marks indicate peak positions in the EDCs. (e)-(h) Zoom-in EDCs at resonance points of VB and CB with fitting peaks appended. (i)-(j) Schematic summary of dispersions upon and at 200 fs with fitting data appended.
	}\label{Fig3}
\end{figure*}

The Floquet origin of this gap is further confirmed by its time evolution, which shows that the gap opening occurs exclusively when the driving field is present and coincides with the emergence of coherent Floquet sidebands. Figures 4(a)-4(e) display second-derivative images at various delay times. At $\Delta t$ = $\pm$ 200 fs, where there is no temporal overlap between the pump and probe pulses, the data show the characteristic gapless Dirac dispersion. Near time zero, however, a pronounced gap is clearly observed in both the VB and CB. The dynamic evolution of this gap is further revealed by EDC analysis in Figs.~4(f)-4(j), where the clear splitting peaks at resonant points [Figs.~4(g)-4(i)] indicates the gap opening. The extracted gap, obtained by fitting energy distribution curves (EDCs) at the resonance points in Fig. 4(k), is plotted as a function of delay time in Fig. 4(l), alongside the fitting curves of the temporal evolution of the Floquet sideband intensity obtained from Fig.~2(n). Both the gap opening and the sideband are observed within a temporal window of 114 fs around time zero, the timescale determined by the pump and probe pulses. This confirms that the light-induced gap and the Floquet sidebands are driven by the coherent light field, establishing their Floquet origin from the time domain perspective.

\begin{figure*}[htbp]
	\centering	
	\includegraphics[width=16.8cm]{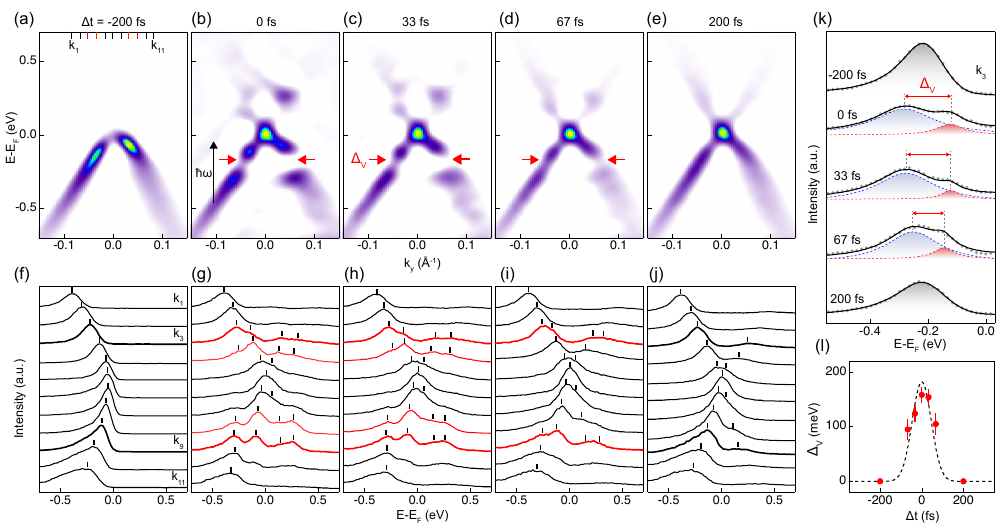}
  \caption*{Fig.~4. Dynamical evolution of the light-induced gap, confirming its Floquet origin. (a)-(e) The second-derivative images measured at different delay times. (f)-(j) EDCs extracted from data in Fig.~2(a) and 2(d)-(g) respectively. Tick marks indicate peak positions in the EDCs. (k) EDCs at resonance points for various delays, with fitted results appended. (l) Extracted Floquet-induced gap at different delay times. The black curve is the same as S3 in Fig.~2(n).
  }\label{Fig4}
\end{figure*}

\section*{Conclusions and Perspectives}

By pumping high-quality graphite with an intense MIR pulse, we provide direct spectroscopic evidence of robust Floquet-induced gaps in both the VB and CB of graphite, despite the presence of photo-excitation and interlayer coupling. The Floquet-induced gap, Floquet sideband and relaxation of photo-excited carriers are disentangled through their distinct timescales. Note that while the carrier relaxation time slightly increases from monolayer graphene to bulk graphite\cite{zhang2025spatial}, carrier relaxation occurs on a timescale of at least several hundred femtoseconds, much longer than the 114 fs temporal window of Floquet-induced gap, allowing the gap to be detected before scattering sets in. Such large timescale mismatch enables the formation of Floquet-induced gap despite the presence of additional scattering channels, and here the short pump and probe pulses are important for the successful detection of such gaps.

The observation of Floquet-induced avoided-crossing gaps in bulk graphite opens opportunities for exploring light-field-engineered phases in graphite. Looking forward, several important questions remain to be addressed both experimentally and theoretically. First of all, regarding the interplay between coherent light-field dressing and scattering following photo-excitation: if a longer pump pulse with strong enough light-field strength is applied, can the light-induced hybridization gap still survive and be detected experimentally? Secondly, for the photon energy used here, the dispersion at the corresponding $k_z$ resembles that of monolayer graphene. What happens if the probe photon energy is tuned to access different $k_z$ values? Will the splitting of the Dirac cone  affect the Floquet-induced gaps? Finally, a central question is what light-induced emergent phenomena can be realized in graphite? How do they differ from those in monolayer graphene, and what exotic phases can be achieved in graphite beyond those predicted in graphene? Addressing these questions will pave the way toward harnessing Floquet engineering to design and control novel quantum phases in bulk materials.

\begin{addendum}

\item [Acknowledgement] 
This work is supported by the National Natural Science Foundation of China (Grant No.~12234011,12421004), Tsinghua University Initiative Scientific Research Program (Grant No.~20251080106), the National Natural Science Foundation of China (Grant No.~52388201, 12327805), and New Cornerstone Science Foundation through the XPLORER PRIZE.

\end{addendum}

\bibliography{reference}

\end{document}